# A new hybrid approach for crude oil price forecasting: Evidence from multi-scale data

Yifan Yang, Ju'e Guo, Shaolong Sun*, Yixin Li

School of Management, Xi'an Jiaotong University, Xi'an, 710049, China

Corresponding author: School of Management, Xi'an Jiaotong University, Xi'an, 710049, China.

Tel.: +86 1591106725.

E-mail address: sunshaolong@xjtu.edu.cn (S.L. Sun).

**Abstract:** Faced with the growing research towards crude oil price fluctuations influential factors following the accelerated development of Internet technology, accessible data such as Google search volume index (GSVI) are increasingly quantified and incorporated into forecasting approaches. In this paper, we apply multi-scale data that including both GSVI data and traditional economic data related to crude oil price as independent variables and propose a new hybrid approach for monthly crude oil price forecasting. This hybrid approach, based on "divide and conquer" strategy, consists of K-means method, kernel principal component analysis (KPCA) and kernel extreme learning machine (KELM), where K-means method is adopted to divide input data into certain clusters, KPCA is applied to reduce dimension, and KELM is employed for final crude oil price forecasting. The empirical result can be analyzed from data and method levels. At the data level, GSVI data perform better than economic data in level forecasting accuracy but with opposite performance in directional forecasting accuracy because of "Herd Behavior", while hybrid data combined their advantages and obtain best forecasting performance in both level and directional accuracy. At the method level, the approaches with K-means perform better than those without K-means, which demonstrates that "divide and conquer" strategy can effectively improve the forecasting performance.

**Keyword:** Crude oil price forecasting, GSVI data, Kernel extreme learning machine, Herd behavior, Divide and conquer

## 1. Introduction

Crude oil, as the blood of the industry, plays an important role in the global economic market, whose price fluctuation has a significant impact on political and economic activities around the world [1,2]. Policymakers who can accurately forecast crude oil price fluctuations can make prospective economic and political policies to gain advantages in a complex international environment. Additionally, since the futures market of crude oil is one of the biggest commodity markets, the availability of accurate forecasting for high profits is one factor that is considered beneficial for both investors and companies. Therefore, based on the conviction that crude oil price forecasting can foster financial market and thus international prestige, crude oil price forecasting has received increased attention, and thus many practical measures and academic research are directed toward improving the accuracy of crude oil price forecasting.

However, accurately forecasting crude oil price is a challenge because of complicated influential factors, such as oil supply and demand, additional commodity market, stock index, world events, etc. [3-7]. Forecasting research is dominated by traditional econometric models and machine learning models. Traditional econometric models include: Random Walk (RW) models [8-10], Autoregressive Integrated Moving Average (ARIMA) models [11,12], Generalized Autoregressive Conditional Heteroscedasticity (GARCH) family models [11-15]. And machine learning models include: Neural network (NN) models [16-19]. Support Vector Machine (SVM) models [20-24], Wavelet-based models [18,19,24,25].

However, both models have their disadvantages. Traditional econometric models have to make assumptions in advance and show poor performance in capturing nonlinear features [26], and machine learning models suffer from overfitting and parameters sensitive problems [27]. It is therefore necessary to propose hybrid approaches to remedy these shortcomings, Yu et al. [26] first applied an "EMD (Empirical Mode Decomposition)-FNN (Feed-forward Neural Network)-ALNN (Adaptive Linear Neural Network)" hybrid approach to forecast daily crude oil price. Jammazi and Aloui [28] applied wavelet decomposition and ANN models to forecast crude oil price. Zhang et al. [20] adopted Ensemble Empirical Mode Decomposition (EEMD) to decompose crude oil price series and forecast the subseries respectively by Least Square Support Vector Machine with Particle Swarm Optimization (LSSVM-PSO) and GARCH model. Tang et al. [29] also utilized EEMD and Random Vector

Functional Link (RVFL) network for crude oil price forecasting. Wang and Wang [30] combined Multi-Layer Perception (MLP) and Elman Recurrent Neural Network (ERNN) based on ANN framework for crude oil price forecasting. These studies all adopted single models as benchmarks and demonstrated that the forecasting accuracy of hybrid approaches was significantly higher than that of single models.

Due to the limitation of measurement and data collection, the role of investor attention on price fluctuation has become a research focus only in recent years [31]. Regarding the effect of investor attention on price fluctuation, Behavioral Finance Theory suggests that the asset price is not only determined by its intrinsic value but also influenced by the investor's behavior. Since human attention is a kind of scarce cognitive resource, investors can only pay attention to the assets they concerned. Therefore, the psychological characteristics such as investor confidence can be revealed by the change of investor attention towards special asset, which can further influence the asset price fluctuation. Although the measurement of investor attention is challengeable, the accelerated development of Internet technology provides feasible solutions to deal with this situation. For example, investors use search engines to search terms about assets they concerned, in such way the search volume is counted and a great number of data are collected, which can generate more objective and available measurements than those of traditional methods [32].

In this paper, we use Google search volume index (GSVI), generated by a public tool (https://trends.google.com/) of Google Inc., as a proxy variable of investor attention for three reasons. First, Google search is the most popular search engine that can offer a huge amount of free and available online data. Second, GSVI consists of normalized structural data range from zero to 100, where zero refers that search volume is below a certain threshold, and 100 refers to a higher limit. Third, since this paper focuses on international crude oil price forecasting instead of Chinese domestic crude oil price forecasting, GSVI is more suitable than other search volume indexes such as Baidu search volume index due to its worldwide adoption.

Prior research finds that GSVI data contributes to analyzing and forecasting various social and economic behaviors. In the field of disease surveillance, Ginsberg et al. [33] used GSVI data to build an influenza epidemic forecasting model, which can forecast the intensity and timing of flu outbreaks one to two weeks in advance. Araz et al. [34] used GSVI data to forecast Influenza-Like-Illness (ILI)-related emergency department visits in Omaha. Song et al. [35] found that there is a significant positive correlation between GSVI of stress and the number of suicide in Korea. In the area of macro-economy, Smith [36] highlighted there is a strong correlation between GSVI related to unemployment and the unemployment rate, and he further added GSVI data in a MIDAS regression framework to forecast unemployment in the UK. Li et at. [37] demonstrated that GSVI data and Consumer Price Index (CPI) officially released by the Statistic Bureau of China have a strong correlation, and the MIDAS forecasting model with GSVI outperforms the benchmarks in the reduction of root mean square error (RMSE) over 30%. Goetz and Knetsch [38] included GSVI data in bridge equation models for German GDP forecasting. Additionally, in the research of tourism management, Bangwayo-Skeete and Skeete [39] investigated GSVI related to hotel and flights in forecasting tourism demand of Caribbean by Autoregressive Mixed-Data Sampling (AR-MIDAS) models. Sun et al. [40] forecast tourism arrivals of Beijing by GSVI and Baidu search volume index (BSVI) data, which outperforms the benchmarks without search engine data. Clark et al. [41] applied GSVI data to forecast the demand for tourism arrivals of U.S. National Parks and get a similar result.

However, using GSVI to forecast crude oil price is still in its infancy. Most research focuses on the relation between GSVI of certain events and crude oil price, or applies GSVI data of a few terms to build forecasting models [42-44]. Han et al. [45] first selected a wider set of GSVI of the terms to forecast weekly crude oil price, but they linearly combine all of GSVI data into a composite index as the independent variable, which might cause omission of certain features.

In a nutshell, we consider a series of GSVI data and economic data as independent data, in which GSVI data represent the impact of micro-individual behaviors and economic data represent the impact of macro-economic variables for crude oil price. We also propose a new hybrid approach, K-means+KPCA+KELM, to forecast monthly crude oil spot price based on above data, where K-means method is adopted to divide input data into certain clusters, KPCA is applied to reduce dimension, which can not only explore more features but also solve the overfitting problems. Besides, KELM is employed for final crude oil price forecasting. Based on "divide and conquer" strategy, our proposed new hybrid approach can obtain a better forecasting performance [46].

The remainder of this paper is organized as follows. The next section introduces the methodologies and the framework of our proposed new hybrid approach. The empirical results are outlined in Section 3. After a discussion of the results, the last section concludes this paper.

## 2. Methodology

In this section, our proposed new hybrid approach and its adopted basic models are introduced.

### 2.1 K-means method

K-means method refers to a method of unsupervised learning algorithm, it can divide similar samples into groups according to their distance. K-means method is the most popular algorithm for data clustering in practice with its superior characteristics such as simple, reasonably scalable and easily understandable.

Given a data set $X = \{x_1, x_2, \cdots, x_N\}$, $x_i \in \Re^d (i = 1, 2, \cdots, N)$. K-means method aims to divide $X$ into k clusters $C = \{C_1, C_2, \cdots, C_k\}$, $\cup C_i = X$, $C_i \cup C_j = \emptyset$ for $(1 \leq i \neq j \leq k)$. After randomly pick up $k$ samples from $X$ as initial centroid vector: $\mu = \{\mu_1, \mu_2, \cdots, \mu_k\}$, $\mu_j \in \Re^d (j = 1, 2, \cdots, k)$. K-means method iterates between the following two steps until the centroid vector no longer changes.

**Step 1**: Data Division. Dividing $X$ into $k$ clusters according to the sample labels determine by its closet centroid.

**Step 2**: Updating Centroid. For each cluster, calculating and updating the clustering centroids: $\mu_i = \frac{1}{|C_i|} \sum_{X \in C_i} X$, where $|C_i|(i = 1, 2, \cdots, k)$ are the sample numbers of current clusters.

This paper considers one issue regarding closeness measuring in **Step 1**. Several different types of distance can be selected based on data features. Euclidean distance, the most popular distance, is the "ordinary" (or straight line) distance between two points in Euclidean space; Cosine distance is more about the difference in direction between two vectors, which is often used in text analysis and sentiment analysis.

Correlation distance is selected to measure closeness in this paper since collected data is multivariable time series and our objective is to group time series based on sequence correlation.

$$dist_{cor}(x_i, \mu_j) = 1 - \frac{(x_i - \bar{\bar{x}}_i)(\mu_j - \bar{\bar{\mu}}_j)'}{\sqrt{(x_i - \bar{\bar{x}}_i)(x_i - \bar{\bar{x}}_i)'}\sqrt{(\mu_j - \bar{\bar{\mu}}_j)(\mu_j - \bar{\bar{\mu}}_j)'}} \quad (1)$$

where $\bar{\bar{x}}_i = \frac{1}{d}\left(\sum_{m=1}^{d} x_{im}\right)\vec{1}_d$, $\bar{\bar{\mu}}_j = \frac{1}{d}\left(\sum_{m=1}^{d} \mu_{jm}\right)\vec{1}_d$ and $\vec{1}_d$ is a row vector of $d$ ones.

**2.2 Kernel Principal Component Analysis**

Kernel Principal Component Analysis (KPCA), first proposed by Schölkopf et al. [47], is an improved feature extraction method for Principal Component Analysis (PCA). The core idea of KPCA (shown in **Fig. 1**) is to map input data to a high dimensional space via a nonlinear kernel function, which generalizes PCA from linear to nonlinear situations.

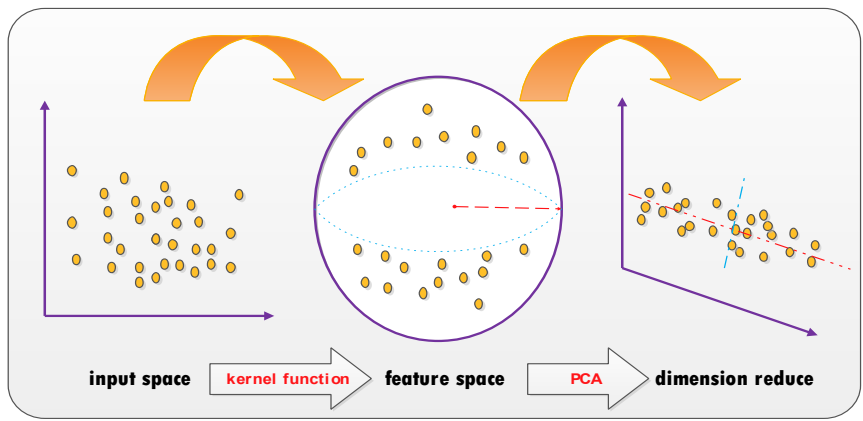

Fig. 1 The basic principle of KPCA

Given a data set $X = \{x_1, x_2, \cdots, x_N\}$, $x_i \in \Re^d$ $(i = 1, 2, \cdots, N)$, and then define an implicit nonlinear map $\varphi(\cdot): x \in \Re^d \to y \in \Re^l$, set the dot product of

mapping samples as the kernel function:

$$k(x_i, x_j) = \varphi(x_i)^T \varphi(x_j) \tag{2}$$

Assuming that the mapping samples are centered:

$$\frac{1}{N} \sum_{i=1}^{N} \varphi(x_i) = 0 \tag{3}$$

Therefore, the covariance matrix is defined as:

$$Q = \sum_{i=1}^{N} \varphi(x_i) \varphi(x_i)^T \tag{4}$$

The following step is to solve the eigenequation:

$$Q\omega_j = \lambda_j \omega_j \tag{5}$$

where $\omega_j \in \Re^I (j = 1, 2, \cdots, I)$ is the eigenvector matrix and $\lambda_j$ is corresponding eigenvalues.

It is known that any vector in a space, even a basis vector, can be represented linearly by all the samples in that space. Thus, the eigenvectors $\omega$ can be written as:

$$\omega_j = \sum_{i=1}^{N} \varphi(x_i) \alpha_{ij} \tag{6}$$

Substituting **Eq. (2)** and **Eq. (6)** into **Eq. (5)**, we get:

$$K\alpha_j = \lambda_j \alpha_j \tag{7}$$

where $K$ is a $N \times N$ kernel matrix with elements $k(x_i, x_j)$:

$$K = \begin{bmatrix} k(x_1, x_1) & k(x_1, x_2) & \cdots & k(x_1, x_N) \\ k(x_2, x_1) & k(x_2, x_2) & \cdots & k(x_2, x_N) \\ \vdots & \vdots & \ddots & \vdots \\ k(x_N, x_1) & k(x_N, x_2) & \cdots & k(x_N, x_N) \end{bmatrix}_{N \times N} \tag{8}$$

Obviously, **Eq. (7)** is a typical eigenvalue decomposition problem, we solve this problem and pick the eigenvectors corresponding to $n'(n' \leq n)$ maximum eigenvalues to combine $n'$-dimensional target feature space.

### 2.3 Kernel Extreme Learning Machine

Extreme learning machine (ELM), one type of effective single-hidden layer feedforward neural network (SLFN), has been widely used in many fields. ELM can generate randomly the weight and bias of a hidden layer, which need not be tuned anymore. After determining the number of hidden nodes and activation function, output weights can be obtained by matrix computations rather than iteration. Therefore, the training speed and generalization ability of ELM are at an advantage, and ELM is thus a significant breakthrough in the development of the traditional neural network.

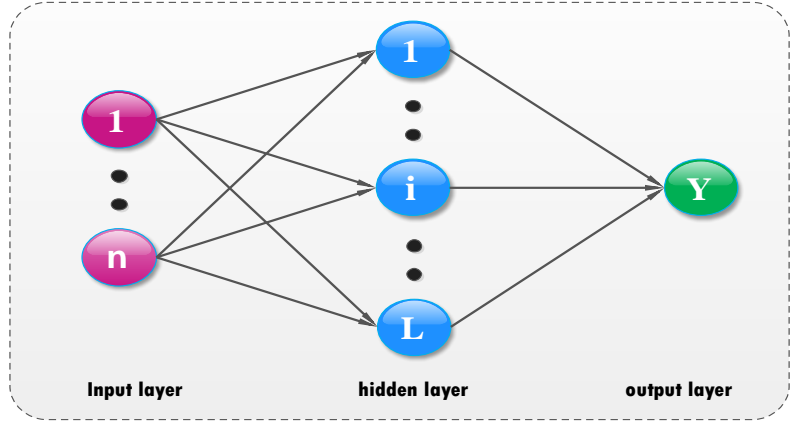

**Fig. 2** The topological structure of ELM

Given N samples $(x_i, y_i)$, $x_i \in \Re^N$, $y_i \in \Re^N$, $i = 1, 2, \cdots, N$, for a typical ELM with n input neurons, L hidden neurons and one output neurons (shown in **Fig. 2**), one can define the output matrix of ELM as:

$$Y = \begin{bmatrix} y_{1j} \\ y_{2j} \\ \vdots \\ y_{mj} \end{bmatrix}_{m \times N} = \begin{bmatrix} \sum_{i=1}^{L} \beta_{i1} h(\omega_i x_j + b_i) \\ \sum_{i=1}^{L} \beta_{i2} h(\omega_i x_j + b_i) \\ \vdots \\ \sum_{i=1}^{L} \beta_{im} h(\omega_i x_j + b_i) \end{bmatrix}_{m \times N}, (j = 1, 2, \cdots, N) \qquad (9)$$

where $\beta$ is the output weights matrix between the hidden layer and the output layer, $h(\cdot)$ is the activation function of the hidden layer, $\omega_i = [\omega_{i1}, \omega_{i2}, \cdots, \omega_{iN}]^T$ is the input weights matrix between i-th input layers and hidden layers, and b is the biases matrix of the hidden layer.

We can also rewrite **Eq. (9)** as:

$$Y = H\beta, Y \in \Re^{N \times m}, \beta \in \Re^{N \times m} \qquad (10)$$

where $H = H(\omega, b) = h(\omega x + b)$ is the output matrix of the hidden layer.

The value of input weights and biases are randomly assigned rather than being tuned. Thus, the output weights are the only unknown parameters, which can be

calculated by the ordinary least square (OLS), the result can be written as:

$$\hat{\beta} = H^{\dagger}Y \tag{11}$$

where $H^{\dagger}$ is denoted as the Moore-Penrose generalized inverse of the output matrix.

Based on Ridge Regression Theory and Karush-Kuhn-Tucker (KKT) theorem, we can also add a positive penalty term $1/C$ to recalculate $\beta$ as:

$$\hat{\beta} = H^T \left( I/C + H H^T \right)^{-1} Y \tag{12}$$

Therefore, the output function of ELM can be presented as:

$$f(x) = H\hat{\beta} = HH^T \left( I/C + H H^T \right)^{-1} Y \tag{13}$$

Although ELM can improve the performance of traditional neuron networks in training speed and generalization ability, it also has disadvantages such as poor robustness. Huang et at. [48] introduced an improved method called kernel-based ELM (KELM) to eliminate these disadvantages. Numerous empirical studies have demonstrated that KELM has a better performance than ELM [49,50]. The main idea of KELM is to replace the activation function of ELM as a kernel function according to Mercer's conditions, the output function of KELM can be presented as:

$$f(x) = H\hat{\beta} = \begin{bmatrix} k(x, x_1) \\ k(x, x_2) \\ \vdots \\ k(x, x_n) \end{bmatrix}^T \left( I/C + H H^T \right)^{-1} Y \tag{14}$$

where $k(x, x_i)$ represents the kernel function.

**2.4 The framework of our proposed new hybrid approach**

In this paper, based on "divide and conquer" strategy, a new hybrid approach named K-means-KPCA-KELM is proposed to forecast crude oil price. It is formulated by following three steps. **Fig. 3** illustrates the framework of our proposed new hybrid approach.

**Step 1**: Data fusion. Collecting the GSVI series of oil-related terms and filter out the irrelevant and unrelated terms, then merge the remaining GSVI series with other economic series as independent variables series.

**Step 2**: Dimension reduction. Applying a K-means method to divide independent variables series into *k* clusters in terms of their correlation degree. For each cluster, KPCA is adopted to reduce data dimensions and obtain low dimension features.

**Step 3**: Forecasting. Combining the above features as the input matrix of KELM to forecast crude oil price.

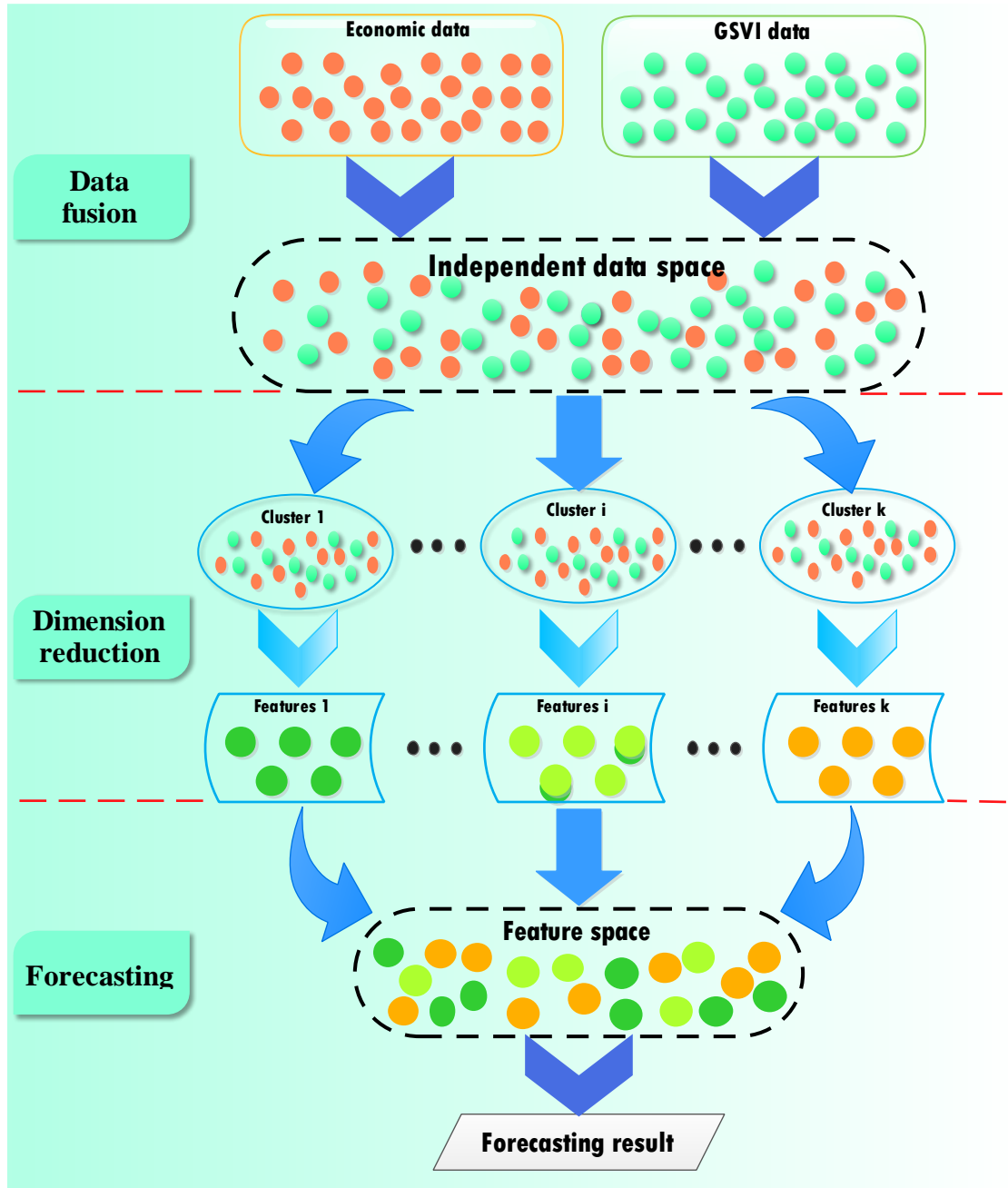

**Fig. 3** The framework of our proposed new hybrid approach

## 3. Empirical study

### 3.1 Data collection

In this paper, the monthly West Texas Intermediate (WTI) crude oil spot price series(shown in **Fig. 4**) extracted from Wind Database (http://www.wind.com.cn/) is used as the dependent variable. In addition, economic dataset and GSVI dataset range from January 2004 to December 2018 are collected as the independent variables. Then we divide those datasets into two parts: the train datasets range from January 2004 to December 2017 and the test datasets range from January 2018 to December 2018. The following subsections describe the above datasets.

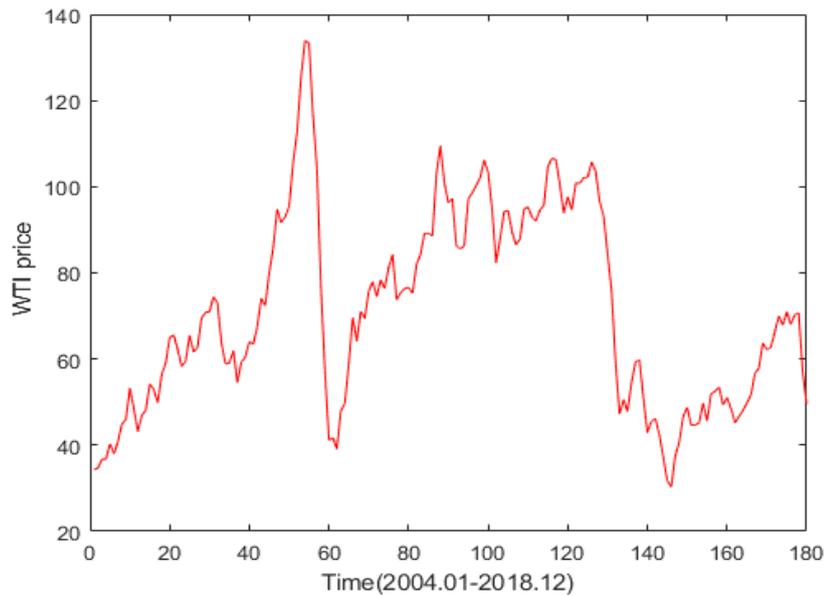

**Fig. 4** Monthly WTI crude oil spot price

**3.1.1 Economic dataset**

Supply, demand, and inventory are called three cornerstones that affect crude oil price [51,52]. Thus, we first consider economic variables related to these basic influencing factors. In this paper, the supply-related variables include crude oil production capacity, refining capacity, consumption structure and replacement cost. The demand-related variables are comprised of the volume of energy consumption and some indexes about global economic development. The inventory-related variables are oil stocks. Moreover, since crude oil price interacts with other economic and financial market activities, relevant variables are added into independent variables as well, such as the monetary market index, commodity market index, and stock market index [53-55]. Economic dataset is described in **Table 1**.

**Table 1** Description of the economic dataset

| First-class index | Second-class index | Variables | Data Source |
|---|---|---|---|
| Supply | Production | Crude Oil Production, Total OPEC | EIA |
| | | Crude Oil Production, Total Non-OPEC | EIA |
| | | Crude Oil Production, World | EIA |
| | Consumption structure | Henry Hub Natural Gas Spot Price | EIA |
| | | Rest US tight oil | EIA |
| | Technology | WTI-Brent spot price spread | EIA |
| | | WTI crack spread: actual value | EIA |
| | | Brent crack spread: actual value | EIA |
| Demand | Consumption | Petroleum Consumption, Total OECD | EIA |
| | | China oil import | Wind database |
| | Global economic development | Fed fund effective | FRB |
| | | Kilian Global economic index | https://sites.google.com/site/lkilian2019 |
| | | US: CPI index: seasonally adjusted | Wind database |
| | | US: CPI: energy: seasonally adjusted | Wind database |
| | | US: PPI: manufacturing sector total | Wind database |
| | | US: PPI: mining sector total | Wind database |
| | | EU 28 Countries: PPI | Wind database |
| | | US PMI index | Wind database |
| Inventory | Inventory | Petroleum Stocks, Total OECD | EIA |
| | | Crude Oil Stocks, Total | EIA |
| | | Crude Oil Stocks, SPR | EIA |
| | | Crude Oil Stocks, Non-SPR | EIA |
| market activity | Monetary Market | Real dollar index: generalized | The federal reserve |
| | | Exchange rate of euro against US dollar | The federal reserve |
| | Stock market | S&P 500 Index | Wind database |
| | | Dow Jones Industrial Index | Wind database |
| | | NYSE Index | Wind database |
| | | AMEX Index | Wind database |
| | | NASDAQ index | Wind database |
| | Commodity market | COMEX: Gold: Future closing price | Wind database |
| | | LME: Copper: Future closing price | Wind database |
| | | Crude oil non-commercial net long ratio | CFCT |

**3.1.2 GSVI dataset**

GSVI data are appropriate proxy variables of investor attention, which can provide numerous information for crude oil price forecasting from a micro perspective. Extracting useful information from a large amount of data is challengeable, we thus apply a three-stage process to refine useful search terms based on these paper [44,45].

Firstly, we build an oil-related terms seed-set based on the following aspects: 1) add the terms directly related to crude oil price such as "oil price", "oil demand" and

"oil supply"; 2) add the terms related to other economic indexes and variables such as "gold price and "GDP"; 3) add oil-related terminologies from the glossary of the Colorado Oil and Gas Conservation Commission (COGCC) and some renewable energy terms; 4) add attention terms with a tendency to fear such as "crisis" and "bankrupt". Secondly, we search the terms of our seed-set in Google Trend and iteratively set recommended terms as second-round search terms. This process is repeated until there are no new terms in the recommended list. Thirdly, we estimate the degrees of relevance with crude oil price series for the above terms by Granger causality test and filter out the terms whose p-value over 0.1. Finally, a set of 40 GAVI terms is built in alphabetical order (shown in **Table 2**).

**Table 2** GSVI data terms

| Alternative energy | Crude oil | Energy security | Great depression | Recession depression |
|---|---|---|---|---|
| Bankrupt | Crude price | Expensive | Greenhouse gases | Recession |
| Bankruptcy | Crude prices | Financial situation | Horizontal drilling | State of the economy |
| Brent crude oil futures | Current interest | Fossil fuel | Interest rates | U.S. economy |
| Brent crude | Economic issues | Frugal | Kerosene | Unemployment |
| Carbon footprint | Economic situation | Gamble | Natural gas price | West Texas Intermediate |
| Carbon intensity | Economy problems | Gas subsidy | Offshore drilling | WTI oil |
| Clean energy | Energy conservation | Going green | Oil price | WTI price |

### 3.2 Performance evaluation criteria

We apply mean absolute percentage error (MAPE), root mean square error (RMSE) and directional accuracy (DA) to evaluate the forecasting accuracy of our proposed new hybrid approach from the level and directional aspect respectively:

$$MAPE = \frac{1}{N} \sum_{t=1}^{N} \left| \frac{y(t) - \hat{y}(t)}{y(t)} \right| \times 100\% \tag{15}$$

$$RMSE = \sqrt{\frac{1}{N} \sum_{t=1}^{N} \left( y(t) - \hat{y}(t) \right)^2} \tag{16}$$

$$DA = \frac{1}{N} \sum_{t=1}^{N} d(t) \times 100\% \tag{17}$$

$$d(t) = \begin{cases} 0 & if \ (y(t+1) - y(t))(\hat{y}(t+1) - y(t)) < 0 \\ 1 & if \ (y(t+1) - y(t))(\hat{y}(t+1) - y(t)) \geq 0 \end{cases} \tag{18}$$

where N denotes the number of observations, $y(t)$ and $\hat{y}(t)$ denote the actual crude oil price and forecasting crude oil price respectively.

MAPE and RMSE measure the level accuracy, the smaller the MAPE/RMSE, the better the level performance. DA measures the directional accuracy, the higher the DA, the better the directional performance.

Moreover, we introduce the improvement rate (IR) to test the superior forecasting ability of our proposed new hybrid approach compared with its benchmarks:

$$IR_{MAPE} = -\frac{MAPE_A - MAPE_B}{MAPE_B} \times 100\% \qquad (19)$$

$$IR_{RMSE} = -\frac{RMSE_A - RMSE_B}{RMSE_B} \times 100\% \qquad (20)$$

$$IR_{DA} = \frac{DA_A - DA_B}{DA_B} \times 100\% \qquad (21)$$

where approach A represents the proposed approach and approach B denotes the benchmark. When approach A outperforms approach B, the value of IR is positive and vice versa.

**3.3 Benchmarks and parameters setting**

To test the superior forecasting ability of our proposed new hybrid approach K-means-KPCA-KELM, six related forecasting approaches are introduced in this subsection as benchmarks. Firstly, we apply three single models for univariate forecasting as follows: one econometric model (ARIMA, one of the most basic models in forecasting field) and two machine learning models (ELM and KELM, which have good performance in crude oil price forecasting).

Then, we adopt four hybrid approaches for multivariable forecasting: K-means +KPCA+ELM, KPCA+KELM, KPCA+ELM and our proposed new hybrid approach, which are used to test the contribution of kernel function and clustering operation for forecasting performance. We apply all the multivariable approaches on three different types of independent variables datasets: GSVI dataset, economic dataset and hybrid dataset (and hybrid dataset includes both GSVI dataset and economic dataset). It is worth mentioning that both K-means-PCA-ELM and PCA-ELM approach have a poor forecasting performance in our datasets. It is suggested that PCA is more suitable for linear problems but crude oil price series is non-linear, uncertain and dynamic, and it is therefore not surprising that PCA is not applied as a dimensional reduce method in this paper.

In this study, the optimum clustering numbers of K-means is determined as 3 according to "Elbow Criterion". The Gaussian kernel function is adopted in both KPCA and KELM as kernel function. The optimal lag of ARIMA is estimated by means of Akaike Information Criterion (AIC) and Schwarz Criterion (SC). The rest of the parameters are selected by trial and error testing by means of the minimization of Mean Absolute Error (MAE).

All models are running by Matlab R2018a software on a server with 4 Core CPU of i5-4590 3.30 GHz, RAM size of 8 GB.

**3.4. Empirical results**

To test our forecasting approach, firstly, three single models are conducted to forecast monthly WTI crude oil spot price in order to find the best single forecasting model. Secondly, four multivariable approaches (including our proposed new hybrid approach) are applied in economic dataset, GSVI dataset and hybrid dataset respectively. The results are interpreted from both data and method perspectives to demonstrate the superior forecasting ability of our proposed new hybrid approach with hybrid dataset. Departing from these, this paper also provides and explains several interesting phenomena at last.

**3.4.1 Forecasting performance comparison of single models**

The forecasting performances of single models with the WTI crude oil spot price dataset are shown in **Fig. 5**, it can be observed that: 1) KELM has the best forecasting performance (MAPE: 8.09%, RMSE: 0.0430, DA: 72.73%), followed by ELM (MAPE: 8.78%, RMSE: 0.0490, DA: 63.64%), and ARIMA ranks the last (MAPE: 12.03%, RMSE: 0.0495, DA: 36.36%). 2) ARIMA, as a traditional econometric model, has poor performance in capturing the nonlinear and dynamic features of crude oil price series, thus the performance of ARIMA is worse than those of the other two machine learning models especially in directional forecasting accuracy. 3) The performance of ELM is quite similar to KELM via both level evaluation criteria and directional evaluation criteria. Therefore, KELM and ELM are considered as the best single models for univariate forecasting and they are selected as the basic models for our hybrid multivariable approaches in the following steps.

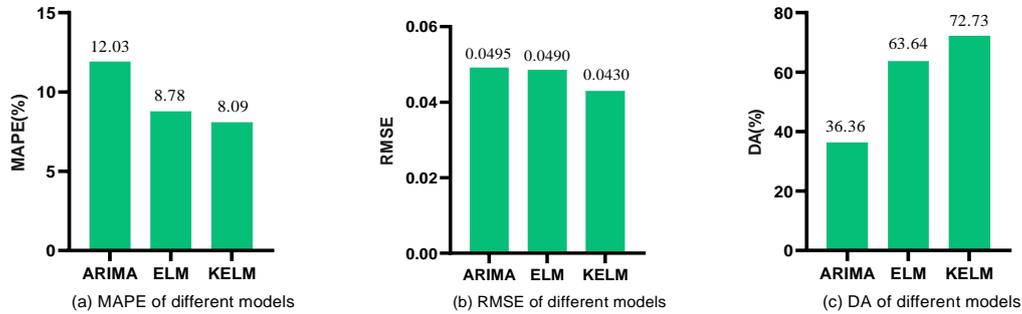

Fig. 5 Forecasting performance comparison of different single models

### 3.4.2 Forecasting performance comparison of multivariable approaches

The forecasting performances of multivariable approaches in three different types of datasets are discussed as follows, **Fig. 6** shows the performance comparison results of four multivariable approaches in different datasets. It is shown that our proposed K-means+KPCA+KELM approach with hybrid dataset has the lowest MAPE: 5.44%, lowest RMSE: 0.0311 and highest DA:90.91%. In general, the multivariable approaches are more efficient than single models. Because the independent variables of multivariable approaches contain a lot of information to capture more features of crude oil price. Moreover, as the results display, the performance of KELM is slightly better than ELM in all groups, it is therefore reasonable to select both ELM and KELM as basic single models.

Next, we apply IR criteria to analyze the empirical results from the data and method perspective respectively, which further support the superior forecasting ability of hybrid dataset and our proposed new hybrid approach.

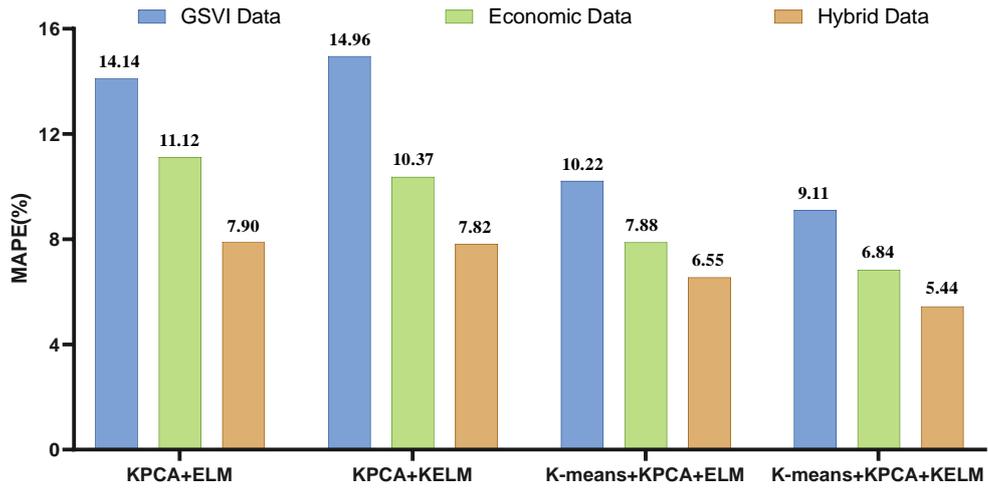

(a) MAPE of different approaches

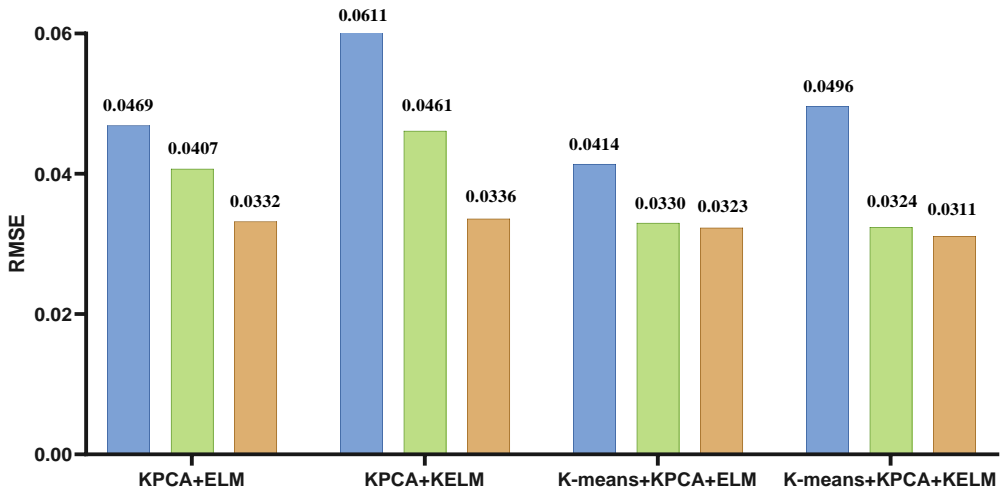

(b) RMSE of different approaches

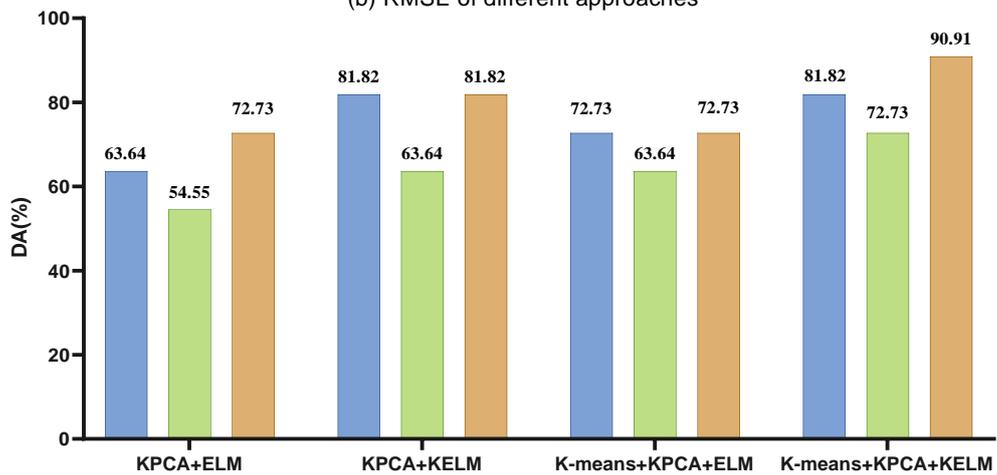

(c) DA of different approaches

**Fig. 6** Performance comparison of different approaches

**Table 3** shows the improvement rate of three evaluation criteria for different datasets, where E, G and H represent economic dataset, GSVI dataset and hybrid dataset respectively. It is clear that: 1) For each group, the IR values of $H \rightarrow E$ and

$H \to G$ are positive in both level and directional performance evaluation criteria, which reveals that hybrid dataset contributes significantly more than only economic dataset or GSVI dataset in crude oil price forecasting. 2) For each group, the IR values of $E \to G$ are positive in level performance evaluation criteria while the values are negative in directional performance evaluation criteria, which shows that economic dataset contributes significantly more than GSVI dataset in level forecasting but with opposite performance in directional forecasting.

**Table 4** displays the contribution of clustering operation in dimension reduction, in which Approach1 and Approach2 represent KPCA+ELM and KPCA+KELM, and Approach3 and Approach4 refer to approaches that combined K-means method with the above models respectively. For each group, Approach3 and Approach4 outperform Approach1 and Approach2 respectively according to the positive IR values. It is obvious that the approaches with K-means method not only obtain the highest level forecasting accuracy (via the positive $IR_{MAPE}$ and $IR_{RMAE}$ criteria) but also acquire the best directional forecasting performance (via the positive $IR_{DA}$ criteria), which indicates that clustering operation contributes a lot for forecasting performance improvements.

**Table 3** IR between different datasets

| Approaches | Datasets | $IR_{MAPE}$ (%) | $IR_{RMSE}$ (%) | $IR_{DA}$ (%) |
| --- | --- | --- | --- | --- |
| KPCA+ELM | $E \to G$ | 21.20 | 13.19 | -14.29 |
|  | $H \to G$ | 44.04 | 29.21 | 14.29 |
|  | $H \to E$ | 28.99 | 18.46 | 33.33 |
| KPCA+KELM | $E \to G$ | 30.65 | 24.50 | -22.22 |
|  | $H \to G$ | 47.70 | 45.01 | 0 |
|  | $H \to E$ | 24.59 | 27.17 | 28.57 |
| K-means+KPCA+ELM | $E \to G$ | 22.89 | 20.19 | -12.5 |
|  | $H \to G$ | 35.85 | 21.87 | 0 |
|  | $H \to E$ | 16.81 | 2.11 | 14.29 |
| K-means+KPCA+KELM | $E \to G$ | 24.92 | 34.74 | -11.11 |
|  | $H \to G$ | 40.32 | 37.21 | 11.11 |
|  | $H \to E$ | 20.51 | 3.79 | 25 |

**Note:** E represents economic dataset; G represents GSVI dataset; H represents hybrid dataset.

Table 4 IR between different approaches

| Datasets | Approaches | IR$_{MAPE}$ (%) | IR$_{RMSE}$ (%) | IR$_{DA}$ (%) |
|---|---|---|---|---|
| GSVI Dataset | $Approach3 \rightarrow Approach1$ | 27.61 | 11.79 | 14.29 |
| | $Approach4 \rightarrow Approach2$ | 39.10 | 18.85 | 0.00 |
| Economic Dataset | $Approach3 \rightarrow Approach1$ | 29.16 | 18.90 | 16.67 |
| | $Approach4 \rightarrow Approach2$ | 34.07 | 29.86 | 14.29 |
| Hybrid Dataset | $Approach3 \rightarrow Approach1$ | 17.02 | 2.64 | 0.00 |
| | $Approach4 \rightarrow Approach2$ | 30.51 | 7.35 | 11.11 |

**Note:** Approach 1 represents KPCA+ELM; Approach 2 represents KPCA+KELM; Approach 3 represents K-means + KPCA + ELM; Approach 4 represents K-means+KPCA+kELM.

### 3.4.3 Discussions

According to the above performance comparisons of both single models and multivariable approaches, it is clear that K-means +KPCA+KELM outperforms other benchmarks in both level and directional accuracy, and hybrid dataset obtains better forecasting performance than single economic dataset or GSVI dataset via both level and directional evaluation criteria. Moreover, we would like to make brief explanations of two interesting phenomena found in **Table 3** and **Table 4**:

(1) As **Table 3** shown, hybrid dataset has the best performance in level accuracy, followed by economic data and GSVI dataset ranks the last, while hybrid dataset has the best performance in directional accuracy, followed by GSVI dataset and economic dataset ranks the last. Economic dataset includes macroeconomic influencing factors and determines the real value of crude oil, which significantly reflect the trend and period components of crude oil price. Besides, economic data is not sensitive to sudden events such as wars and abnormal climate that lead to the short-term fluctuations of crude oil price. GSVI dataset is composed of the proxy variables of investor attention so it can reveal the investor behavior. In the capital market, a single investor always acts according to the actions of other similar investors, buys when others buy, and sells when others sell, this phenomenon is called "Herd Behavior". Thus, investors are sensitive to sudden events and react quickly, thus GSVI dataset has a strong ability to capture the short-term fluctuation components of crude oil price, which contributes a lot for directional forecasting ability. It is nevertheless true that investors tend to exaggerate the degree of crude oil price fluctuations because of "Herd Behavior", which reduces the level forecasting accuracy. However, hybrid dataset, combined economic dataset with GSVI dataset, can not only capture the trend, period components, but also capture the short-term fluctuations components of crude oil price without exaggeration. In brief, economic dataset tends to improve level forecasting accuracy while GSVI dataset tends to improve directional forecasting accuracy, and hybrid dataset combines their advantages to get the best performance in both level and directional forecasting accuracy.

(2) As **Table 4** shown, the approaches with K-means method (K-means +KPCA+KELM and K-means +KPCA+ELM) perform better than corresponding approaches without K-means (KPCA+KELM and KPCA+ELM). Based on "divide and conquer" strategy, our proposed new hybrid approach first divides the input data into k clusters, then individually reduce dimensions for each cluster, and thirdly group these low dimension features as new input data for the forecasting model. Compared with the direct dimension reduction method, "divide and conquer" strategy is more refined and effective, which can discover the unique properties for different components of origin series.

## 4 Conclusions

In this paper, we combined economic dataset with GSVI dataset as independent variables for crude oil price forecasting, where the two datasets reflect the impact of macro-economic variables and micro-individual behavior respectively. In order to fully exploit and utilize the information of above dataset, we proposed a new hybrid approach combined with K-means, KPCA and KELM, where K-means method is applied to divide independent variables into k clusters according to their correlation degree, KPCA is adopted to map independent variables into low dimensional space, KELM is employed for final crude oil price forecasting. Our empirical results show that our proposed new hybrid model significantly outperforms other benchmarks in both level and directional accuracy for each dataset and hybrid dataset performs better than other datasets in both level and directional accuracy for every approach.

Based on these results, the contribution of our work is threefold. Firstly, compared with the traditional econometric model, our proposed single model KELM has a strong ability in capturing the nonlinear and dynamic features of crude oil price and outperform other single models. Secondly, GSVI dataset has a strong ability to capture the short-term fluctuations components of crude oil price, due to the existence of the "Herd Behavior", GSVI dataset often exaggerate the degree of those fluctuations, while economic dataset reflects more trend and period components of crude oil price and less short-term fluctuations components. Our proposed hybrid dataset, composed by economic dataset and GSVI dataset, combines their advantages to capture trend and period as well as short-terms fluctuations components of crude oil price. Thirdly, based on "divide and conquer" strategy, this paper performs a clustering operation before reduce dimension, which prefer to discover more information about crude oil price fluctuations and improve forecasting accuracy.

It is suggested that our proposed new hybrid approach based on "divide and conquer" strategy and multi-scale data fusion especially GSVI data can be applied as independent variables to obtain a better forecasting performance in other complex forecasting issues such as electric power load or consumption forecasting, traffic flow forecasting and PM2.5 concentration forecasting.

However, since this paper applies the most common Gaussian functions as the kernel function in KPCA and KELM, it is suggested that other alternatives functions substitute for Gaussian can further improve the forecasting accuracy. In addition, some parameters in this paper are determined by trial and error testing, which is time-consuming and not suitable for large-scale data processing. Hence, a more appropriate and time-saving method to select optimal parameters should be exercised in future research.

## Conflict of Interests

The authors declare that there is no conflict of interests regarding the publication of this paper.

## Acknowledgment

This research is supported by the National Natural Science Foundation of China (Project No: 71774130).